\documentclass[journal=jacsat,manuscript=article]{achemso}

\usepackage[version=3]{mhchem} 
\usepackage{xcolor}
\usepackage{xr}
\usepackage{soul}
\usepackage{graphicx}

\author{Xiaohe Lei}
\email{xiaohelei@ucsb.edu}
\affiliation{Department of Chemistry and Biochemistry, University of California, Santa Barbara, Santa Barbara, California 93106, United States}

\author{Vojtech Vlcek}
\affiliation{Department of Chemistry and Biochemistry, University of California, Santa Barbara, Santa Barbara, California 93106, United States}
\alsoaffiliation{Department of Materials, University of California, Santa Barbara, Santa Barbara, California 93106, United States}

\title[An \textsf{achemso} demo]
  {Impact of Subsurface Oxygen on CO$_2$ Charging Energy Changes in Cu Surfaces}

\begin{document}


\begin{abstract}
Subsurface oxygen in oxide-derived copper catalysts significantly influences \ce{CO_2} activation. However, its effect on the molecular charging process, the key to forming the \ce{CO_2^{\delta-}} intermediate, remains poorly understood. We employ many-body perturbation theory to investigate the impact of the structural factors induced by the subsurface oxygen on charged activation of \ce{CO_2}. By computing the molecular single-particle state energy of the electron-accepting orbital ($\sigma*$) on Cu (111) surface, we examined how this molecular quasi-particle (QP) energy changes with varied vicinity of adsorption and multiple subsurface oxygen configuration. We demonstrate that subsurface oxygen impairs \ce{CO_2} charging, with its presence and density being influential factors. The non-local potential proves substantial for accurate excitation energy predictions yet is not sensitive to minor atomic structural changes. More importantly, state delocalization and hybridization are critical for determining QP energy. These insights are enlightening for designing atomic architectures to optimize catalytic performance on modified surfaces.
\end{abstract}

Metallic catalysts are widely used for the chemical conversion of \ce{CO_2} as a carbon feedstock to produce valuable hydrocarbon products through catalytic reactions.\cite{liu2012co2,kwak2013heterogeneous,zhang2014competition,matsubu2015isolated,kumar2016new,wang2017tuning,lu2018high,zhang2020co2,franco2020transition,rotundo2021electrochemical,vijay2021unified,hori2008electrochemical,raciti2015highly,eilert2016formation,rhimi2024cu,li2012co2,chi2014morphology,kas2014electrochemical,li2014electroreduction,roberts2015high,ren2015selective,ma2015selective,dutta2016morphology,liu2021co2,ali2022activity} Copper uniquely stands out among transition metals for its ability to efficiently convert \ce{CO_2} into high-value multi-carbon products such as ethylene.\cite{hori2008electrochemical,raciti2015highly,eilert2016formation,rhimi2024cu,li2012co2,chi2014morphology,kas2014electrochemical,li2014electroreduction,roberts2015high,ren2015selective,ma2015selective,dutta2016morphology,liu2021co2,ali2022activity} Oxide-derived Cu has been observed to significantly enhance both activity and selectivity towards high-value $C_{2+}$ products in the \ce{CO_2} reduction reaction (CO2RR).\cite{li2012co2,chi2014morphology,kas2014electrochemical,li2014electroreduction,roberts2015high,ren2015selective,ma2015selective,dutta2016morphology,liu2021co2,ali2022activity} Recent studies have revealed that subsurface oxygen plays a critical role in \ce{CO_2} activation, which is the rate-determining step in CO2RR.\cite{favaro2017subsurface}

Experimental studies have revealed a substantial and steady presence of subsurface oxygen during CO2RR, at least in the initial stages of the reduction reaction.\cite{favaro2017subsurface,cavalca2017nature,eilert2017subsurface,wang2022direct} Given the observed promotion of CO2RR by subsurface oxygen\cite{favaro2017subsurface,cavalca2017nature,eilert2017subsurface,liu2021co2,wang2022direct}, understanding how it modifies the Cu substrate and interacts with adsorbed molecules has become an intriguing research topic. Density functional theory (DFT)-based studies have demonstrated significant stabilization of \ce{CO_2} adsorption due to subsurface oxygen.\cite{fields2018role,ye2019dramatic,liu2021co2} Moreover, subsurface oxygen has been shown to contribute to strengthened \ce{CO} binding to the surface, which is a critical intermediate product in CO2RR following \ce{CO_2} activation.\cite{eilert2017subsurface,cavalca2017nature,liu2017stability,fields2018role} However, the impact of subsurface oxygen on charge carrier behavior, which describes the key electronic excitation event in the elementary reaction step, has not been thoroughly explored.\cite{hori2008electrochemical,chernyshova2018origin,chernyshova2019activation}

It is well established that while multiple reaction pathways for CO2RR exist on metallic surfaces\cite{ye2019dramatic}, the rate-determining step is the first step in CO2RR, involving the activation of \ce{CO_2} on the surface\cite{hori2008electrochemical,chernyshova2018origin,chernyshova2019activation},
\begin{equation}
    \ce{CO_2} + e^- + * \rightarrow ^*\ce{CO_2^-}
\end{equation}
In this reaction, the \ce{CO_2} molecule gains an additional electron, forming a bent structure and coordinating with the surface (denoted by $*$). The energy scale of this additional electron directly informs us about the propensity to form such a negative ion intermediate (which may be transient) and thus provides insight into how this behavior changes in the presence of subsurface oxygen. 
Focusing on this vertical electronic excitation, we aim to investigate whether the charge injection propensity is altered by subsurface oxygen and how the electronic structure changes. This investigation should provide valuable knowledge about the impact of subsurface oxygen at the single-electron scale.

Kohn Sham (KS) DFT has been the most widely used quantum mechanical framework for studying electronic structures in large-scale electron systems. It provides reliable ground state charge density, even for complex interface systems.\cite{biller2011electronic,egger2015reliable,liu2017energy,wu2021identification,bhandari2021achieving} However, its efficacy in characterizing individual electronic levels is limited, as the KS single particles are fictitious, non-interacting entities, and their associated eigenvalues do not correspond to physical observables.\cite{martin2016interacting} To capture the single-particle state more accurately, we employ many-body perturbation theory (MBPT) using the quasiparticle (QP) picture for excited single-particle states, which directly determines the excitation energy. 

In this context, we initiate our investigation by utilizing the KS density computed via DFT and apply GW approximation (GWA) to address and analyze the molecular QP state more comprehensively. The self-energy $\Sigma$ incorporating the many-body interaction potential is computed as the product of Green's function $G$ and the screened coulomb interaction $W$ in the time domain, $\Sigma = iGW$. We employ a stochastic GW package that utilizes stochastic vector techniques to reduce computational costs\cite{neuhauser2014breaking,vlvcek2018swift}. 
A one-shot $G_0W_0$ calculation should sufficiently estimate the self-energy $\Sigma$.\cite{giantomassi2011electronic} Both $G_0$ and $W_0$ are constructed using the KS eigenstates and KS Hamiltonian.
This approach has demonstrated significant improvements in accurately predicting experimental band gaps and molecular ionization potentials, establishing it as a reliable theoretical method for analyzing molecule-metal interfaces.\cite{rignanese2001quasiparticle,tamblyn2011electronic,kaplan2015off,vlcek2017stochastic,li2022benchmark} 

In this study, we utilize the QP picture to elucidate the impact of subsurface oxygen on the charged excitation event in \ce{CO_2} activation. We examine multiple structural factors, including the depth and density of oxygen, surface atom displacement induced by oxygen, and adsorption proximity to oxygen. Our findings reveal that the presence of oxygen negatively affects the electron injection process to the molecule. Interestingly, the QP energy level shows little sensitivity to structural changes in the metal substrate; instead, the density of oxygen emerges as a more influential factor. Furthermore, we emphasize the critical roles of non-local potential and state hybridization in accurately capturing and understanding the QP state. This comprehensive approach provides valuable insights into the complex interplay between the structural changed induced by subsurface oxygen and the adsorbate electronic excitation in molecule-metal interface systems.

In accordance with the proposed mechanism of \ce{CO_2} activation on the metal surface, electrons are injected into the $\sigma*$ anti-bond of the adsorbed \ce{CO_2}.The charge injection into the $\sigma*$ anti-bonding orbital corresponds to the formation of the activated $CO_2^\delta-$ species on the metal surface, i.e., the charged excitation of the molecule. The $\sigma*$ anti-bonding orbital is computed as the lowest unoccupied molecular orbital (LUMO) in the $CO_2$ molecule in DFT. Consequently, this study focuses on the LUMO-like QP of \ce{CO_2} in the CO$_2$-Cu(O) interface system. This examination sheds light on the variations in the charge injection propensity as well as the non-local correlation effects across different substrates with subsurface oxygen.

Before proceeding to the MBPT investigation of the electronic structure properties, we first establish the optimized atomic structure of the interface system through DFT. It is widely recognized that the embedded oxygen often triggers substantial rearrangements among surface metal atoms.\cite{cavalca2017nature,liu2017stability} This phenomenon introduces additional variables, including surface roughness, into consideration. Our objective is to create a model closely resembling a \ce{CuO_{sub}} surface in comparison to a pristine Cu surface. This aims to elucidate the effects resulting from the introduction of subsurface oxygen. Consequently, we exclusively examine scenarios where oxygen atoms are positioned within the interstitial spaces beneath the top one or two layer of Cu atoms, ensuring minimal geometric influence. The detailed atomic structure parameters and optimization process are included in the Supplementary Information (SI). 

To ensure a realistic slab model, we calculate the Fermi level of Cu (111) slab by incrementally adding layers of Cu, and we observe convergence (with a Fermi level difference of less than $50$ meV) when using four monolayers of copper. Thereby a Cu (111) slab with four Cu monolayers is used in the subsequent study. To accommodate multiple adsorption sites in different lateral distances to the oxygen, a larger surface area of 4$\times$4 unit cells is used. In case the work function of the substrate is size sensitive\cite{schulte1976theory} thus giving rise to the unnecessary fluctuation of the computed electronic levels, in this work we keep the size of the substrate of (4$\times$4) unit cells $\times$4 layers. 

Next, we compute the KS orbitals along with their KS eigenvalues for the interface system which is subsequently used as the starting point for the stochastic GW calculation. Also we obtain the LUMO for the molecule in the vacuum by KS DFT. These calculations employ the same norm-conserved pseudopotential with the PBE functional as in structure optimization above. In principle, the molecular orbital particularly the frontier ones like HOMO and LUMO, would significantly hybridize with the substrate states resulting in energy level realignment and orbital broadening. Herein we identify the most LUMO-like canonical KS state as the representative molecular charge carrier state, which is identified with the most overlap ($p = \langle \phi^{KS}| LUMO \rangle$) with the fully localized LUMO in the isolated molecule. 

The self-energy and the QP energy for these charge carrier states are then computed employing stochastic $G_0W_0$ approach. The QP energy is found through the fixed point equation,
\begin{equation}
\epsilon_{QP} = \epsilon_{KS}+\langle\phi|\hat{\Sigma}_P(\omega=\epsilon_{QP})+\hat{\Sigma}_X-\hat{v}_{xc}|\phi\rangle
\end{equation}
where $\epsilon_{KS}$ is the KS eigenvalue, $\hat{v}_{xc}$ is the approximated exchange-correlation potential (PBE functional in this study) and $\hat{\Sigma}_X$ is the static exchange self-energy and $\hat{\Sigma}_P$ is the dynamical self-energy that encompasses induced density-density interactions, incorporating the plasmonic response of the substrate. A chemical accuracy of $75$ meV, which corresponds to a $20$ times difference of charged excitation propensity in Boltzmann factor ($e^{-\frac{\epsilon}{k_BT}}$)at room temperature ($300$ K), is used for the statistic error convergence. 

We first inspect the pristine Cu slab as a baseline. The Fermi energy for the larger-area pristine 4-layer Cu (111) surface is $-5.21$ eV, which agrees reasonably well with the experimental value of the work function $-4.94$ eV for the Cu (111) single crystal presuming the vacuum level is around zero.\cite{gartland1972photoelectric} Introducing subsurface oxygen into the same size of the substrate causes a change of $<60$ meV in the Fermi energy of the metal substrate. This variation is smaller than the $75$ meV statistical error convergence used in the QP energy calculation throughout this paper. 

\begin{figure}[htbp]
        \centering
        \includegraphics[width=0.5\textwidth]{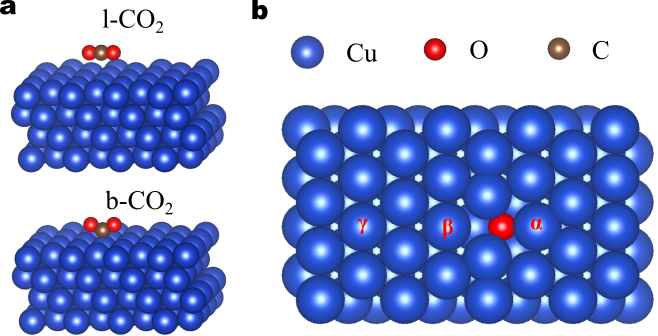}
        \caption{\textbf{(a)} \ce{CO_2} is adsorbed atop an atomic site on a $4\times4$ unit-cell $\times4$ layers - Cu (111) surface slab. Upper: physisorption scenario with a linear \ce{CO_2} probe on the Cu (111) surface. Bottom: chemisorption scenario with a bent \ce{CO_2} probe on the Cu (111) surface. \textbf{(b)} The adsorption position scheme relative to the oxygen position. In the case with two oxygen embedded in the substrate, there would be only $\alpha$ and $\beta$ positions available.} 
        \label{fig1}
    \end{figure}

In the following we first examine the physisorption scenario for a linear \ce{CO_2}. In practice, we select three adsorption sites $\alpha$, $\beta$ and $\gamma$ (Fig.\ref{fig1}.b) ordered by the increasing distance from the oxygen incorporated to the substrate, and four distinct types of oxygen incorporation into the substrate to study the effect of the depth and coverage (i.e., the density of oxygen surface incorporation to the subsurface) as shown in Fig.S4: Substrate A serves as a reference Cu slab with 1/16 MLE coverage of subsurface oxygen located between the top two layers, $0.36$ \AA\, beneath the Cu surface, and \ce{Cu3} $0.13$ \AA\, above the surface; Substrate B has deeper oxygen situated between the top two layers, $0.83$ \AA\, beneath the Cu surface, and \ce{Cu3} $0.20$ \AA\, above the surface; Substrate C has oxygen between the second and third layers, representing the deepest oxygen case without direct contact with the surface Cu atoms; Substrate D represents a system with higher oxygen coverage of 1/8 MLE with the same \ce{Cu3O} geometry used in substrate A. Finally, a model chemisorption scenario is examined to reveal the impact of oxygen on the bonded adsorbate. All the previously mentioned adsorption sites and substrate types are inspected in this context as well. We will describe both scenarios independently in two subsections.

\begin{figure}[htbp]
        \centering
        \includegraphics[width=0.8\textwidth]{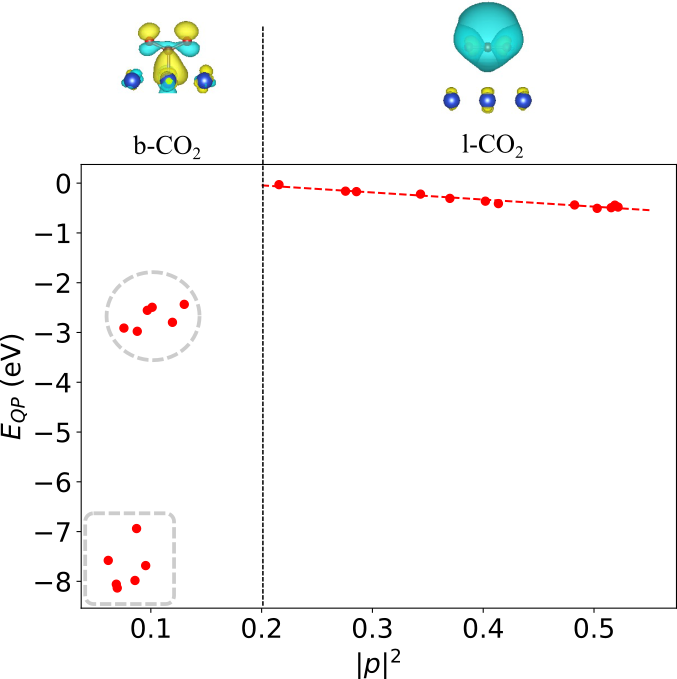}
        \caption{Plot of QP energies of the LUMO, illustrating both physisorption and chemisorption scenarios as a function of the $\sigma*$ fraction $|p|^2$ in the QP state. Above the plot, spatial visualizations of the chemisorbed (b-\ce{CO2}) and physisorbed (l-\ce{CO2}) LUMO are shown. Physisorbed cases distribute in the stronger $\sigma*$ character region (right), while chemisorbed cases cluster in the weaker $\sigma*$ character region (left). For physisorption, QP energy exhibits a linear trend with $|p|^2$, fitted by $E_{QP} = -1.42*|p|^2 + 0.24$ ($R^2=0.955$). In the chemisorption scenario, electron states are marked by a dashed grey circle, and hole states by a dashed grey square.} 
        \label{fig3}
    \end{figure}

Before turning to the detailed results for the physisorbed scenario, we inspect how the LUMO state and its character changes in individual cases studied. In fact, we observe that though the identified LUMO resembles a $\sigma*$ (Fig.\ref{fig3}) the corresponding $|p|^2$ value which naturally signals the $\sigma*$ orbital character strength, typically ranges from $0.20$ to $0.55$. The state is thus strongly hybridized with surface states. The most LUMO-like canonical KS state exhibits $|p|^2 =0.52$ for the \ce{CO2-Cu} system without any oxygen doping. While the near higher energy level like $\pi*$ (namely LUMO+1 and LUMO+2) might contribute to the state mixing, in this case, their contribution is only $0.2\%$, indicating negligible molecular orbital characters other than LUMO. However, with oxygen doping, the contribution from the $\pi*$ anti-bonding orbital is significantly higher, reaching up to $23\%$.

Previously, we found that a more hybridized single-particle state, i.e., low $|p|^2$, should have an energy level closer to the substrate's Fermi level.\cite{lei2023exceptional} As we shall demonstrate here, the state mixing changes this picture substantially and the LUMO QP here. Yet, in neither case we found a more complicated state reordering that would place the molecular LUMO+1 or LUMO+2 states energetically closer to the Fermi level compared to the state that most resembles the gas-phase LUMO orbital. The lack of state reordering is illustrated in Fig.S6. As we illustrate at the end of this subsection, the contribution fractions from $\pi*$ and $\sigma*$ orbitals are negatively correlated, with the $\sigma*$ character remaining dominant in the selected LUMO.

The LUMO QP energy level is found consistently roughly $5$~eV above the Fermi energy of Cu surface, which is a high barrier for the direct electron injection from the substrate. The LUMO QP in pristine case is at $-0.49$ eV. We observe a clear impact on the single-particle energy level depending on the proximity of the molecular adsorption site to the oxygen, yet, the injection barrier is not \textit{decreased} by the presence of subsurface oxygen in either case studied. The most influenced is the nearest adsorption position $\alpha$, at which the LUMO QP state is destabilized (i.e., shifted up in energy) by approximately $300$ meV on average between $\alpha^A$ and $\alpha^B$ cases and compared to the pristine case. A higher density of oxygen has an even stronger impact, destabilizing the LUMO QP level by over $450$ meV ($\alpha^D$). Note that such discrepancy is not observed in KS picture, where the averaged KS eigenvalues at $\alpha$ position differ from the pristine case by less than $40$ meV, within chemical accuracy.

The $\beta$ position is the ``nearest neighbor'' adsorption distance to the subsurface oxygen position (i.e., just one site away from $\alpha$) and surprisingly appears to be the least impacted, maintaining the energy level around $-0.46$~eV. This value is practically equivalent to the pristine case and underscores abrupt variations of the QP injection barrier energies as a function of their position. In contrast, the ``next-nearest neighbor'' position, $\gamma$, is impacted by the subsurface oxygen, with the LUMO QP level being approximately $290$ meV higher on average than in the pristine case. A direct relation between the local chemical environment caused by the adsorption proximity and the QP energy is not obvious here. Later we see this factor of the local chemical environment is taken account in the hybridization of the QP state which dominates the QP energy level ordering.

The QP energy correction (i.e., the difference between QP and the underlying KS energies) depends on the magnitude of the self-energy representing the many-body interactions and it is on average $50\%$ higher in magnitude compared to the pristine case. This already indicates a significant impact of subsurface oxygen to the molecular LUMO QP energy level in terms of the substrate renormalization effect. Yet, these effects overall do not lead to stabilization of the LUMO level in either case.

    \begin{figure}[htbp]
        \centering
        \includegraphics[width=0.8\textwidth]{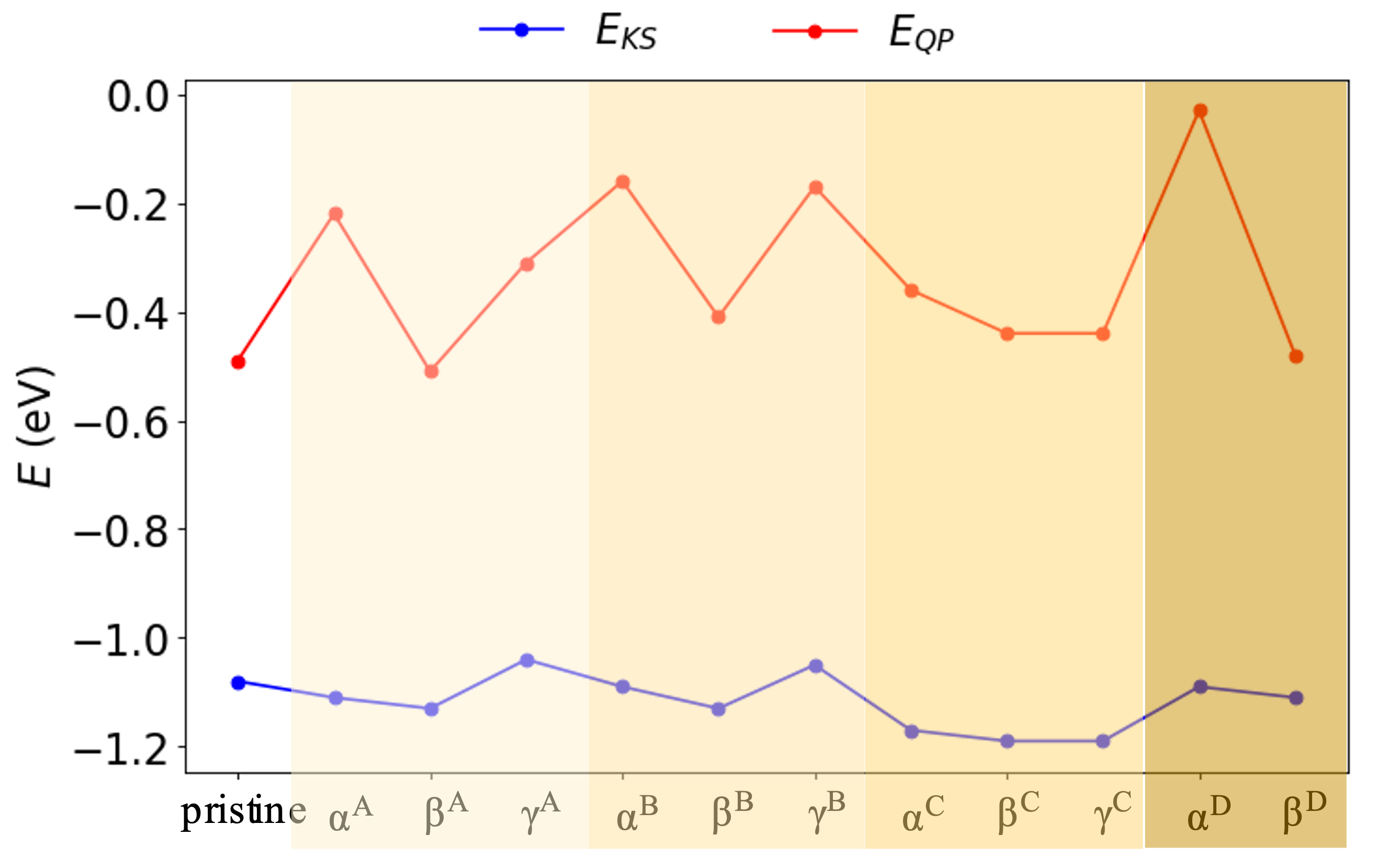}
        \caption{LUMO energies in the physisorption scenario. Red dots represent QP energies predicted by $G_0W_0$ calculations, while blue dots show KS eigenvalues from KS DFT (PBE).} 
        \label{fig2}
    \end{figure}

Besides the lateral adsorption distance, the depth of oxygen plays a non-negligible role in impacting the molecular energy level as well. The relaxed Cu structure already is indicative that such an effect is nontrivial because ``bumping'' is seen when oxygen is incorporated between the top two Cu layers. The shallower oxygen case (A) has the LUMO QP energy level lower than the deeper oxygen (B) by up to $260$ meV. For the even deeper oxygen position in the second sublayer (C), we see that the surface doping impact on the QP state level is further lessened. $\alpha^A$ and $\alpha^B$ have the LUMO QP level on average $300$ meV higher than in the pristine case, LUMO QP for $\alpha^C$ is $-0.36$ eV, which is $230$ meV higher than in the pristine case. As for the distant adsorption cases, The LUMO QP levels remain almost identical for $\beta^C$ position and $\gamma^C$, around $-0.44$ eV which is within the chemical accuracy deviation from the pristine case. This indicates that in such type of subsurface oxygen substrate without direct surface atom contact with the oxygen, the induced local chemical environmental changes from oxygen are presumably minor when the adsorption position is laterally further from the oxygen. Hence, despite we observe that non-local interactions have a pronounced impact on the sites neighboring the oxygen dopant, such an observation does not translate uniformly to all the sites. 

As mentioned, the subsurface oxygen between the top two layers induces surface atom displacement by around $0.1$ \AA. To disentangle the role of the the structural changes alone we investigated two cases of substrate A: one with oxygen removed (but keeping the atomic rearrangement in place) and another with surface deformation eliminated (but keeping the oxygen incorporated). As shown in Fig. S6, such changes in chemical environment merely alter the LUMO QP level at $\beta$ and $\gamma$ positions within a reasonable range ($<100$ meV). Surprisingly, at the $\alpha$ position, the case without surface deformation does not exhibit significant changes in the degree of hybridization (manifested by the practically identical $|p|^2$ value compared to the original case substrate A). In the \textit{absence} of oxygen, the $|p|^2$ value of the state increases by $0.16$ and the LUMO QP energy level stabilizes by $370$ meV. This indicates that the impact on the QP energy level is primarily due to the presence of subsurface oxygen rather than surface deformation.

Beneath all these structural factors, the degree of state mixing emerges as the underlying dominant factor influencing the QP energy level.
Fig.\ref{fig3} demonstrates that more LUMO-like QP orbital, which has a smaller $1-|p|^2$ value, is counterintuitively associated with lower QP energy that is closer to the Fermi level. This is because the molecular $\pi*$ orbitals (LUMO+1 and LUMO+2), which are energetically near LUMO but higher, contribute to the state mixing though $\sigma*$ still contribute higher fraction than $\pi*$. As shown in Fig.S5, the fraction of these higher-energy unoccupied molecular levels is strongly correlated with seemingly exponential dependence with the fraction of LUMO $|p|^2$. As mentioned at the beginning of this section, the projection onto $\sigma*$ in the pristine case is $0.52$, and it is associated with barely $0.2\%$ fraction of $\pi*$ orbitals. In contrast, a QP state in the subsurface oxygen scenario with a $0.28$ LUMO contribution could have a $\pi*$ fraction as large as $0.23$. With a smaller $|p|^2$ value for the LUMO, the state incorporates significantly more higher-energy state character, which raises the QP energy level, moving it further away from the Fermi level. 

Now we comment on the role of the many-body effects. The QP energy correction based on KS eigenvalues, $\Sigma - v_{xc}$, is always positive for these injected electron states, destabilizing the LUMO QP level compared to the KS DFT prediction. Both the non-local self-energy and the semi-local approximated exchange-correlation potential show a linear trend with the strength of the LUMO character, $|p|^2$. As shown in Fig.S9, with stronger LUMO character, the semi-local approximation $v_{xc}$ approaches the quantity of many-body interaction $\Sigma$ computed in $G_0W_0$. Zooming into the self-energy components, both exchange and polarization contribute to stabilizing the QP state. In practice, the exchange contribution is consistently around $1.5\times$ larger than the polarization term. Unlike for the occupied states, where the exchange contribution is typically larger by almost one order of magnitude\cite{lei2023exceptional}, the empty molecular states are significantly more affected by the non-local correlation fluctuations on the metallic substrate. Yet these effects are much less influenced by the degree of hybridization (Fig.S9) as the dynamical correlations are dominated by the surface states which are only minimally altered by small oxygen doping.

Finally, we turn our attention to the chemisorption scenario to examine the impact of subsurface oxygen on the bonded adsorbate. The real-space visualization of LUMO of the isolated molecule also exhibits a ``bent" configuration mirroring its molecular structure described in the method section. As expected, the QP state shows significantly more hybridized compared to the physisorption cases. This is evidenced in the generally weaker character of $\sigma*$, with its fraction consistently below $0.15$ as opposed to the least value of $0.2$ for the physisorbed case. On the other hand, the analysis of the $\pi*$ fraction reveals that these higher energy levels mix in at approximately $1\%$ on average and never exceed $3\%$, i.e., significantly less than in the physisorbed case. These observations indicate that the orbital contribution from the substrate is substantially higher than that in the physisorption scenario.
Notably, the $\sigma*$ anti-bonding orbital now displays a large lobe along the bonding axis in the direction of contact with the surface (Fig.\ref{fig3}), consistent with bonding with the surface and distinctively more hybridized single electron states. 

In examining the possible state reordering between $\pi*$ and $\sigma*$ orbitals, we found that while LUMO+1 remains above LUMO, LUMO+2 now sinks deeply below the Fermi level (Fig.S11). The corresponding $|p|^2$ values of these LUMO+2 orbitals exceed $0.25$, indicating a significant $\pi*$ orbital character. In the pristine case, the LUMO+2 QP is at $-18.2$ eV, demonstrating remarkable stabilization.
For such deep valence states, the influence of subsurface oxygen is minimal, with changes in QP energy less than $2.5\%$ with the presence of subsurface oxygen. Nevertheless, the $\alpha$ site shows a deviation from the pristine case $1.5 \times$ larger than at other sites. This observation also suggests that the impact of subsurface oxygen is restricted.

Furthermore, we observe that in the real-space representation, the molecular $\sigma*$ orbital non-negligibly projects onto the occupied subspace in the interface system. The occupied fraction, $|p|^2 = \sum_i^{occ}|\langle \phi_i^{KS}| \varphi^{LUMO} \rangle|^2$, ranges between $0.50-0.62$ when the molecule is chemisorbed on the surface (while it is between $0.20$ and $0.26$ in the physisorption scenario).
This substantial increase strongly indicates the formation of activated \ce{CO_2}$^{\delta-}$ species on the surface with the molecular reactive anti-bonding orbital spatially half filled with the charge, aligning well with previously reported charge redistribution analysis\cite{chernyshova2019activation}. Indeed, as we show below, in some instances, the QP state falls below the Fermi level and hence it becomes part of the occupied subspace. Given the distinctive nature of these two situations - one where the QP state remains above the Fermi level and another where it falls below - we will discuss them separately.

\begin{figure}[htbp]
        \centering
        \includegraphics[width=0.8\textwidth]{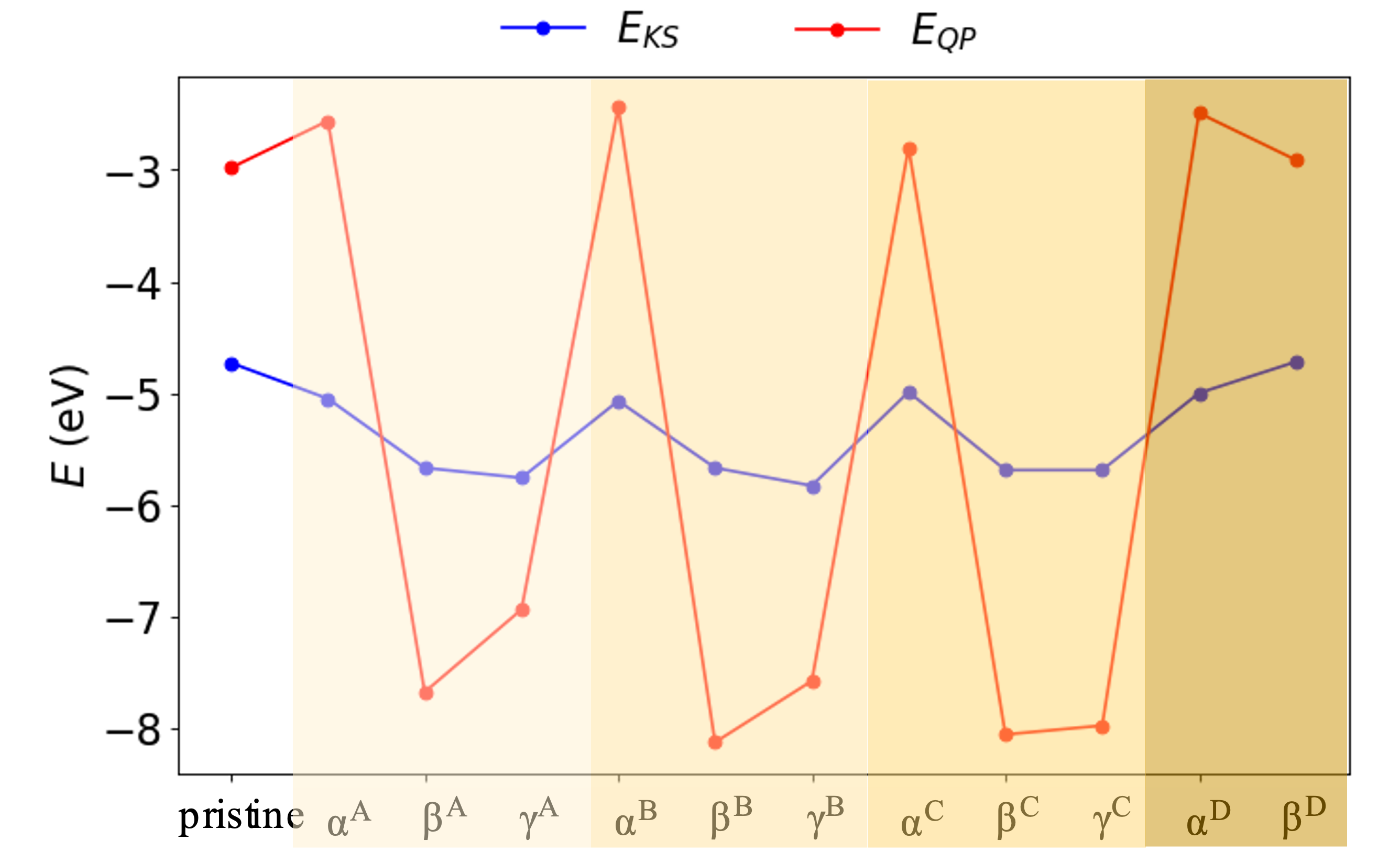}
        \caption{LUMO energies in the chemisorption scenario. Red dots represent QP energies predicted by $G_0W_0$ calculations, while blue dots show KS eigenvalues from KS DFT (PBE).} 
        \label{fig4}
    \end{figure}

In the pristine case, the LUMO QP state is at $-2.98$ eV, more than $2$ eV higher than the Fermi level. Although this is significantly closer to the metal substrate's Fermi level compared to the physisorption scenario, it still represents a large energy gap for direct charge injection from the substrate to the adsorbate. This implies that even after \ce{CO2} undergoes a chemical transformation from its inert linear geometry to a bent activated form, direct charge injection from the substrate remains relatively unfavorable. For cases with subsurface oxygen, the predicted injected electron state energy levels ($\alpha^A$, $\alpha^B$, $\alpha^C$, $\alpha^D$ and $\beta^D$) are, on average, $340$ meV higher than in the pristine case. Yet again, the QP states are thus destabilized by the presence of subsurface oxygen. It's important to note that KS eigenvalues predict the LUMO near resonance with the Fermi level for all chemisorption cases, as shown in Fig.\ref{fig4}. However, QP predictions with MBPT change this picture substantially. 

For $\alpha^A$, $\alpha^B$ and $\alpha^D$ cases, where the molecule directly adsorbs on the Cu atom bonded with the oxygen atom, our results reveal an intriguing discrepancy between QP calculations and KS-DFT predictions regarding the effect of subsurface oxygen on the LUMO state. While KS-DFT suggests that subsurface oxygen stabilizes the LUMO by over $250$ meV, QP calculations show a destabilization of the LUMO in the presence of subsurface oxygen as mentioned earlier. The stabilization phenomenon exhibited by KS prediction is aligned with the local electronic effect induced by the oxygen group. The subsurface oxygen group exerts an electron-withdrawing effect, making the bonded surface Cu atom less electron-rich. This results in smaller electrostatic repulsion to electron transfer in its bonding area, potentially explaining the KS prediction of a lower electron injection energy barrier in the presence of subsurface oxygen. While the destabilization effect of the subsurface oxygen exhibited in the QP energy would need to include the consideration of the state hybridization effect. With oxygen present, the QP state exhibits stronger $\sigma*$ character, with $|p|^2$ values on average $27\%$ higher than in the pristine case. This indicates that the QP state has a smaller fraction of the substrate and is thus less delocalized within it, hindering the charge redistribution from the substrate to the molecule in the subsurface oxygen case. The QP energy results demonstrate that the predicted electron injection energy barrier is higher in the subsurface oxygen cases. This suggests that the enhanced state delocalization introduced by oxygen outweighs the classical electrostatic effect. This finding underscores the importance of advanced computational methods in capturing subtle electronic effects that may be overlooked by standard DFT calculations.

Regarding indirect adsorption positions, only in the denser oxygen substrate (D) does the $\beta$ QP remain above the Fermi level. (Fig.\ref{fig4}) In other substrates, the QP energy levels sink more than $2$ eV below the Fermi level. These states exhibit even greater hybridization, with $|p|^2$ values consistently below $0.1$, exhibiting a very weak $\sigma*$ character. Their KS eigenvalues closely resemble those observed in the injected electron cases, near the Fermi level ranging from $-5.65$ to $-5.85$ eV. However, the QP energy levels for these hole states are far low beneath the Fermi level, ranging from $-8.13$ to $-6.94$ eV, making these hole states distinctively different from the electron states observed in the direct adsorption cases in the energy scale. This is due to the fact that incorporating many-body electron-electron interaction in $G_0W_0$ calculation leads to a significant reduction in energy levels for occupied states and an increase for unoccupied states.

To summarize, in examining the reactive molecular QP state of the physisorbed adsorbate, we found that subsurface oxygen hinders rather than promotes the charged excitation event. The structural influence induced by atomic position rearrangements on the QP energy level is minimal. However, the presence of oxygen is crucial in changing the energy scale of the molecular QP state, causing up to approximately $0.5$ eV discrepancy in QP energy level. The oxygen density is found to be more influential than atomic position changes.

Besides these structural factors, the molecular orbital contribution $|p|^2$, an indicator of state hybridization, emerges as the determining factor for the QP energy level. While we intuitively expected QP states with less $\sigma*$ fraction to be closer to the Fermi level due to more substrate state hybridization, in reality, the QP energy level trends upward with less $\sigma*$ fraction. This is explained by the negative correlation between the $\sigma*$ fraction and the higher-energy orbital $\pi*$ fraction in the QP state. With more hybridization of these higher energy levels in the QP state and less hybridization of LUMO, the overall QP state is destabilized.

Further inspection of potential terms highlights the necessity of considering non-local effects encapsulated in the self-energy term when determining single-particle state energy levels. Yet, we again remark that these non-local effects are ``short-sighted" and significantly affect mostly the sites in the vicinity of the (sub)surface dopant. These potentials also exhibit a linear trend along the $|p|^2$ value. These findings collectively suggest that state hybridization $|p|^2$ could be a good parameter for predicting charged excitation propensity.

In the inspection of the chemisorbed adsorbate, we found the $\sigma*$ orbital is even more hybridized with the substrate, leaving the most featured LUMO QP with a molecular orbital character of less than 0.15. QP energy calculations show that such b-\ce{CO_2} is effectively activated with its most featured LUMO+2 occupied and spatially localized LUMO half-filled. Notably, QP energy predicts that subsurface oxygen should destabilize the hole states, contrasting with the stabilization effect suggested by KS DFT calculations. This draws our attention to the delocalization of quantum states beyond the local classic electrostatic effect induced by the electron-withdrawing nature of the oxygen atom.

These findings have significant implications for understanding and potentially controlling \ce{CO2} activation on modified copper surfaces. The complex interplay between subsurface oxygen, adsorption geometry, and electronic structure suggests that careful tuning of surface properties could optimize \ce{CO2} activation for applications such as electrochemical reduction.\cite{matsubu2015isolated,kumar2016new,wang2017tuning,cheng2023restructuring} Furthermore, our results underscore the importance of employing advanced computational methods like $G_0W_0$ in conjunction with DFT to accurately predict and interpret the electronic properties of adsorbates in heterogeneous catalysis. Additionally, experimental validation of these theoretical predictions would be valuable in further refining our understanding of \ce{CO2} activation on modified copper surfaces.

\begin{acknowledgement}
This work was supported by the NSF CAREER award (DMR-1945098). Use was made of computational facilities purchased with funds from the National Science Foundation (CNS-1725797) and administered by the Center for Scientific Computing (CSC). The CSC is supported by the California NanoSystems Institute and the Materials Research Science and Engineering Center (MRSEC; NSF DMR 2308708) at UC Santa Barbara.
\end{acknowledgement}

\begin{suppinfo}
Optimize atomic structure; b-\ce{CO_2} adsorption geometry; Substrate parameters; Simulation cell parameters; Supplementary plots: physisorption; Supplementary plots: chemisorption.
\end{suppinfo}

\bibliography{subo}
\end{document}